\documentclass[11pt]{article}

\title{\bf New Explicit Binary Constant Weight Codes from Reed-Solomon Codes}

\author{Liqing Xu and Hao Chen
  \thanks{L. Xu and H. Chen are with the School of Sciences, Hangzhou Dianzi University, Hangzhou  310018, Zhejiang Province, China. e-mail: lqxu@hdu.edu.cn, haochen@hdu.edu.cn. This research  was supported by NSFC Grant 11371138.}}

\begin{document}

\maketitle

\begin{abstract}

\end{abstract}
Binary constant weight codes have important applications and have been studied for many years.  Optimal or near-optimal  binary 
constant weight codes of small lengths have been determined. In this paper we propose  a new  construction of explicit binary constant weight codes  from  $q$-ary Reed-Solomon codes. Some of our binary constant weight codes are optimal or new.  In particular new binary constant weight codes $A(64, 10, 8) \geq 4108$ and $A(64, 12, 8) \geq 522$ are constructed. We also give explicitly constructed binary constant weight codes which improve  Gilbert and Graham-Sloane lower bounds in some range of parameters. An extension to algebraic geometric codes is also presented.


\section{Introduction}

 A binary contant weight $(n, d, w)$ code is a set of vectors in ${\bf F}_2^n$ such that\\

1) every codeword is a vector of Hamming weight $w$;\\

2) the Hamming distance $wt({\bf x}-{\bf y})$ of any two codewords ${\bf x}$ and ${\bf y}$ is at least $d$.\\

Binary constant weight codes have important applications (\cite{CHT, Massey, Immink}). In coding theory to determine the maximal possible size $A(n,d,w)$ for a binary constant weight $(n,d,w)$ code is a classical problem which has been studied by many authors (\cite{Johnson, Graham,XingLing,Vardy, Brouwer, Chee, EtzionVardy,table}). For these low $d$ and $w$ and lengths $n \leq 65$ or $n \leq 78$, the previous best known lower bound for $A(n, d, w)$ has been given in \cite{table}. For the upper bounds of $A(n, d, w)$ we refer to the Johnson bound (\cite{Johnson,EtzionVardy}).\\

{\bf Johnson upper bound.} {\em If $n \geq w >0$ then 􏰉
$A(n,d,w) \leq[ \frac{n}{w}A(n-1,d,w-1)]$ and $A(n, d, w) \leq [\frac{n}{n-w} A(n-1, d, w)]$.}\\

The following lower bounds are the most known lower bounds for binary constant weight codes (\cite{Graham}).\\

{\bf Gilbert type lower bound.} {\em $A(n, 2d, w) \geq \frac{\displaystyle{n \choose w}}{\Sigma_{i=0}^{d-1} \displaystyle{w \choose i} \cdot {n-w \choose i}}$.}\\

{\bf Graham-Sloane lower bound.} {\em Let $q$ be the smallest prime power satisfying $q \geq n$ then $A(n, 2d, w) \geq \frac{1}{q^{d-1}} \displaystyle{n \choose w}$.}\\

However the binary constant weight codes in the Gilbert type lower bound is not constructed and the argument is only an existence proof. The binary constant weight codes in the Graham-Sloane lower bound were not  explicitly given, since one has to search at least $q^d$  codes to find the desired one (see \cite{Graham}, page 38). The Graham-Sloane lower bound was improved in \cite{XingLing} by the using of algebraic function fields. However the binary constant weight codes in \cite{XingLing} were not explicitly given, since the construction there was a generalization of \cite{Graham}.\\

In this paper we propose a general construction of  explicit binary constant weight codes  from general $q$-ary Reed-Solomon codes. This is a strict improvement on the previous constructions in \cite{DeVore,Ericson}. Many of our constructed binary constant weight codes have nice parameters. Some of them are new or optimal. In particular two new better binary constant weight codes are presented. We also give an extension to algebraic geometric codes.

\section{Constant weight codes from  Reed-Solomon codes}

In this section we give explicit binary codes from  Reed-Solomon codes. The first step  construction Proposition 2.1-2.2 is  the same as the DeVore's about  restricted isometry matrices in the compressed sensing (\cite{DeVore}). Actually it is also another form of Ericson-Zinoviev construction in \cite{Ericson}. The main results Theorem 2.1-2.2 are strcitly better than these previous constructions.\\

\subsection{Construction}

{\bf Proposition 2.1.} {\em $A(q^2, 2q+2-2r, q) \geq q^r$ if $r-1<q$.}\\

{\bf Proof.} For a polynomial $f$ with degree less than or equal to  $r-1$ in ${\bf F}_a[x]$, we get a length $q^2$ vector $v_f=(f_{(a,b)}) \in {\bf F}_2^{q^2}$. It  is determined by its $q^2$ coordinates $f_{(a,b)}$ for $(a, b) \in {\bf F}_q \times {\bf F}_q$. Here $f_{(a,b)}=0$ if $f(a) \neq b$, $f_{(a,b)}=1$ if $f(a)=b$. Then we have $q^r$ such length $q^2$ codewords from all degree $\leq r-1$ polynomials, each of these codewords has weight $q$. For any two such codewords from polynomials $f$ and $g$, the intersection of their supports are exactly the points $(x, f(x)=g(x))$. Since there are at most $r-1$ zeros of the polynomial $f(x)-g(x)$ we get the conclusion.\\

{\bf Proposition  2.2.} {\em For any positive integer $w \leq q$ where $q$ is a prime power we have $A(wq, 2w+2-2r, w) \geq q^r$ if $q \geq w>r-1$ is satisfied.}\\

{\bf Proof.} We use $q^r$  functions restriced to a subset ${\bf W}$ in ${\bf F}_q$ satisfying $|{\bf W}|=w$.\\

However this explaination of codewords as support functions naturally leads us to add some new codewords. This will give us new better binary constant weight codes which have never been found before.\\

{\bf Theorem 2.1.} {\em $A(q^2, 2q+2-2r, q) \geq q^r+q$ if $2 \leq r<q+1$. $A(wq, 2w+2-2r, w) \geq q^r+w$ if $q \geq w >r-1$ satisfied.}\\

{\bf Proof.} We add these weight $w$ codewords supported at the positions ${u} \times {\bf W}$ where $u$ can be any element in the set ${\bf W}$. Since the supports of these codewords of weight $w$ are disjoint and the support of each such codeword has only one common position with the support of each codeword  $v_f$, the conclusion is proved.\\

{\bf Theorem 2.2.} {\em If $r \geq 3$, $A(wq, 2w+2-2r, w) \geq q^r+w+\frac{w}{[\frac{w-r+1}{q-w+r-1}]+1}$ if $r-1 <  w \leq q$ satisfied.}\\

{\bf Proof.} We add the codewords supported at the following sets. Set ${\bf W}=\{P_1,...,P_w\}$. Every support set is included in at most two $P_i \times {\bf F}_q$'s and the intersection of two such support sets have at most $r-1$ elements. In $P_1 \times {\bf F}_q$ we take $w$-element subset and there are $q-w$ elements remained. In $P_2 \times {\bf F}_q$ we only need to take $w-(q-w+r-1)$-element subset and there are $2(q-w+r-1)$ elements remained. At the $j$-th step we have $(j-1)(q-w+r-1)$ elements remained in the set $P_j \times {\bf F}_q$. \\

When $(j-1)(q-w+r-1) \geq w-r+1$, we add $j+1$ such codewords satisfying that\\
1) The intersection of any two of their supports has at most $r-1$ elements;\\
2) The intersection of any such support with the any image support has at most $2\leq r-1$ elements.\\

Thus finally we add at least $w+\frac{w}{[\frac{w-r+1}{q-w+r-1}]+1}$ such weight $w$ codewords.\\

In a  recent paper \cite{EtzionVardy} of T. Etizon and A. Vardy constructed binary constant weight codes by using the constant dimensional subspace codes. They proved  $A(2^{2m-1}, 2^{m+1}-4, 2^m) = 2^{2m-1}+2^{m-1}$. Our this lower bound $A(2^{2m}, 2^{m+1}-4, 2^m) \geq 2^{3m}+2^m$ can be compared with their result.\\

From Theorem 2.1-2.2  the following binary constant weight codes can be constructed, which can be compared with the best known ones in \cite{table}. The code $A(64, 10, 8) \geq 4108$ and $A(64, 12, 8) \geq 522$ are new and better than the previously best known ones. Many other codes attain the best known lower bounds or optimal values.\\

\begin{center}
{\bf Table 1} Explicit constant weight codes from RS\\
\bigskip
\begin{tabular}{||c|c||}\hline
Explicit codes& lower bound and upper bound in \cite{table} \\ \hline
$A(25, 8, 5)=30$&30\\ \hline
$A(35, 8, 5)=54$&56\\ \hline
$A(40, 8, 5)=69$&72 \\ \hline
$A(42, 10, 6) =55$&55-56 \\ \hline
$A(48, 10, 6)=70$&72 \\ \hline
$A(49, 12, 7) \geq 56$& 56\\ \hline
$A(49, 10, 7) \geq 350$&385-504\\ \hline
$A(64, 10, 8) \geq 4108$&4096--8928\\ \hline
$A(64, 12, 8) \geq 522$&520-720\\ \hline
$A(56, 12, 7) \geq 71$&71-72\\ \hline
$A(56, 10, 7) \geq 519$&583-728\\ \hline
$A(81, 16, 9) \geq 90$&90\\ \hline
$A(64, 14, 8) \geq 72$&72\\ \hline
$A(63, 12, 7) \geq 88$&88--90\\ \hline
$A(63, 10, 7) \geq 736$&831--1116\\ \hline
$A(66, 10, 6) \geq 127$&143 \\ \hline
$A(72, 14, 8)  \geq 89$&89--90\\ \hline
$A(77, 12, 7) \geq 128$&no\\ \hline
\end{tabular}
\end{center}

In the following table 2 we give some small binary constant weight codes from  Theorem 2.1- 2.2, which are compared with the closest codes in \cite{table},  the Gilbert and Graham-Sloane lower bounds. There is no entry in the previous table \cite{table} for these parameters.\\

\begin{center}
{\bf Table 2} Explicit constant weight codes from RS\\
\bigskip
\begin{tabular}{||c|c||c||c||}\hline
Explicit codes& closest codes in \cite{table}&GS bound &G bound\\ \hline
$A(88, 10, 8) \geq 14657$&$A(64, 10, 8) \geq 4096$&1071.8&556.99\\ \hline
$A(72, 10, 8) \geq 6573$&$A(64, 10, 8) \geq 4096$&445.4&255.39\\ \hline
$A(88, 14, 8) \geq 133$&$A(72, 14, 8)\geq 89$&$\leq1$&6.51 \\ \hline
$A(99, 16, 9) \geq 133$&$A(81, 16, 9) \geq 90$& $\leq1$&5.29 \\ \hline
$A(110, 18, 10) \geq 133$&$A(91, 18, 10) \geq 91$&$\leq1$&4.44 \\ \hline
\end{tabular}
\end{center}

\subsection{Explicit binary constant weight codes improving Gilbert and Graham-Sloane lower bounds}

From the Graham-Sloane lower bound $A(q^2, 2(q+1-r), q) \geq \frac{1}{q^{2(q-r)}} \displaystyle{q^2 \choose q}$. With the help from the Sterling formula $lim \frac{n!}{\sqrt{2\pi n}(\frac{n}{e})^n}=1$. We get $\displaystyle{q^2 \choose q} \approx \frac{(eq)^q}{\sqrt{2\pi(q-1)}}$. Thus the Graham-Sloane lower bound in this case $n=q^2$ and $w=q$ is $A(q^2, 2(q+1-r), q) \geq \frac{e^q q^{2r-q}}{\sqrt{2\pi(q-1)}}$. On the other hand the binary constant weight codes staisfying $A(q^2, 2(q+1-2), q) \geq q^r$ are explicitly given in Proposition 2.1. When $r=cq$, where $c$ is a positive constant $0<c<1$, it is clear our lower bound is much better than the Graham-Sloane bound when $q$ is very large. \\

From a simple computation we get $\frac{\displaystyle{q^2 \choose q}}{\Sigma_{i=0}^{q-r} \displaystyle{q \choose i} \cdot {q^2-q \choose i}} \leq  \frac{\displaystyle{q^2 \choose q}}{\displaystyle{q \choose q-r} \cdot \displaystyle{q^2-q \choose q-r}}  \leq  \frac{\displaystyle{q^2 \choose q}}{\displaystyle{q \choose q-r} \cdot (q-1)^{q-r}}$.  From the Sterling formula if $\frac{e^q}{\displaystyle{q \choose r}} <1$ our explicit binary constant weight codes in Proposition  2.1 improve the Gilbert type bound. When $r=cp$,  $c$ is a positive constant very close to $1$,  it is clear $\frac{e^q}{\displaystyle{q \choose r}} <1$ is valid. \\

In \cite{Ericson,XingLing} binary constant weight codes from algebraic curves are used to improve Gilbert and Graham-Sloane lower bounds. However their codes cannot be explicitly constructed. Our codes in Proposition 2.1 are explicitly constructed. As far as our knowledge this is the first improvement on these two lower bounds by explicit constructed binary constant weight codes.\\

\section{Extension to algebraic geometric codes}

\subsection{First step construction}

{\bf Proposition  3.1.} {\em  Let  ${\bf X}$ be a projective non-singular algebraic curve defined over a finite field ${\bf F}_q$ of genus $g$,   ${\bf P}=\{P_1,...,P_{|{\bf P}|}\}$ be a set of ${\bf F}_q$ rational points on the curve ${\bf X}$ and ${\bf G}$ be a ${\bf F}_q$ rational divisor satisfying $deg {\bf G} \geq 2g-1$. Then we have $A(q|{\bf P}|, 2|{\bf P}|-2deg {\bf G}, |{\bf P}|) \geq q^{deg {\bf G}-g+1}$.}\\

{\bf Proof.} For each $f \in L({\bf G})$, a  length $q |{\bf P}|$ vector ${\bf v}_f=(f_{(a,b)}) \in {\bf F}_2^{q|{\bf P}|}$,  where $(a,b) \in {\bf F}_q \times {\bf P}$,  is defined as follows. $f_{(a,b)}$ is $0$ if $f(b) \neq a$ and $f_{a,b)}$ is $1$ if $f(b)=a$. We have $dim(L({\bf G}))=deg{\bf G}-g+1$ if $deg{\bf G} \geq 2g-1$.  There are at least $q^{deg{\bf G}-g+1}$ such codewords.  On the other hand the intersection of two supports of two such codewords associated with functions $f$ and $g$ are exactly these positions $(x, f(x)=g(x))$. Thus it is the zero locus of the function $f-g \in L({\bf G})$. There are at most $deg{\bf G}$ common positions at the intersection of supports of two such codewords.\\

The above construction can be generalized to higher dimension case. Let ${\bf Y}$ be a non-singular algebraic projective manifold defined over ${\bf F}_q$. The set of all ${\bf F}_q$ rational points of this manifold is denoted by ${\bf Y(F_q)}$. For an effective divisor ${\bf D}$ on ${\bf Y}$, we will use  the function space $L({\bf D})$ which consists of all rational functions on ${\bf Y}$ with poles at most $-{\bf D}$ (\cite{Hartshorne}).  In many cases the dimension of this function space can be computed from the Riemann-Roch theorem (\cite{Hartshorne}). For any rational function $f \in L({\bf D})$,  a length $q \cdot |{\bf Y(F_q)}-{\bf D}|$ codeword ${\bf v(f)}_h \in {\bf F}_2^{q|{\bf Y(F_q)}-{\bf D}|}$, where $h=(a,b)$, $ b \in {\bf Y(F_q)}-{\bf D}$ and $a \in {\bf F}_q$, is defined as follows. ${\bf v(f)}_h$ is zero if $f(b) \neq a$,  and ${\bf v(f)}_h$ is $1$  if  $f(b)=a$. Thus the Hamming weight of this codeword is exactly $|{\bf Y(F_q)}-{\bf D}|$. The cardinality of the intersection of the supports of two such codewords ${\bf v(f_1)}_h$ and ${\bf v(f_2)}_h$ is at most the number of zero points in ${\bf Y(F_q)}$ of the function $f_1-f_2$. That is, the number of common positions in the supports of two such codewords is equal to or smaller than the maximal possible number of ${\bf F}_q$ rational points on members of the linear system $Linear({\bf D})$. We denote this number by $N({\bf D})$. \\

{\bf Proposition 3.2.} {\em We have a $A(q \cdot |{\bf Y(F_q)}-{\bf D}|, 2(|{\bf Y(F_q)}-{\bf D}|-N({\bf D})), |{\bf Y(F_q)}-{\bf D}|) \geq  q^{dim(L({\bf D}))}$.}\\

This part of the construction can be seen as a direct application of \cite{Ericson} to algebraic geometric codes.\\

\subsection{Adding new codewords}

{\bf Curve case:} We add the codewords supported at the following sets. Set ${\bf P}=\{P_1,...,P_{|{\bf P}|}\}$. Every support set is included in at most several $P_i \times {\bf F}_q$'s. the intersection of two such support sets have at most $deg {\bf G} $ elements. Suppose $|{\bf P}|=rq+r'$ where $r' \geq 0$. \\

In $P_1 \times {\bf F}_q, ..., P_{r+1} \times {\bf F}_q$ we take $|{\bf P}|$-element subset and there are $q-r'$ elements remained. In $P_{r+2} \times {\bf F}_q,...,P_{2r+2} \times {\bf F}_q$ we only need to take $|{\bf P}|-(q-r'+deg{\bf G})$-element subset and there are $2(q-r'+deg{\bf G})$ elements remained. At the $j$-th step we have $(j-1)(q-r'+deg{\bf G})$ elements remained in the set $P_{j(r+1)} \times {\bf F}_q$. \\

When $(j-1)(q-r'+deg{\bf G}) \geq |{\bf P}|-deg{\bf G}$, we add $j+1$ such codewords satisfying that\\
1) The intersection of any two of their supports has at most $deg{\bf G}$ elements;\\
2) The intersection of any such support with the any image support has at most $r+1$ elements.\\

If $r+1 \leq deg{\bf G}$, finally we add at least $[\frac{|{\bf P}|}{r+1}]+\frac{|{\bf P}|}{[\frac{|{\bf P}|-deg{\bf G}}{q-r'+deg{\bf G}}]+1}$ such weight $|{\bf P}|$ codewords.\\

{\bf Theorem 3.1.} {\em  Let  ${\bf X}$ be a projective non-singular algebraic curve defined over a finite field ${\bf F}_q$ of genus $g$,   ${\bf P}=\{P_1,...,P_{|{\bf P}|}\}$ be a set of ${\bf F}_q$ rational points on the curve ${\bf X}$ and ${\bf G}$ be a ${\bf F}_q$ rational divisor satisfying $deg {\bf G} \geq 2g-1$. Suppose $|{\bf P}|=rq+r'$ and $q>r' \geq 0$. We assume $r+1 \leq deg{\bf G}$. Then we have $A(q|{\bf P}|, 2|{\bf P}|-2deg {\bf G}, |{\bf P}|) \geq q^{deg {\bf G}-g+1}+[\frac{|{\bf P}|}{r+1}]+\frac{|{\bf P}|}{[\frac{{\bf P}|-deg{\bf G}}{q-r'+deg{\bf G}}]+1}$.}\\

{\bf Higher dimension case:} We add the codewords supported at the following sets. Set ${\bf Y(F_q)-D}=\{P_1,...,P_N\}$. Every support set is included in at most several $P_i \times {\bf F}_q$'s. the intersection of two such support sets have at most $N({\bf D}) $ elements. Suppose $N=rq+r'$ where $r' \geq 0$. \\

In $P_1 \times {\bf F}_q, ..., P_{r+1} \times {\bf F}_q$ we take $|{\bf P}|$-element subset and there are $q-r'$ elements remained. In $P_{r+2} \times {\bf F}_q,...,P_{2r+2} \times {\bf F}_q$ we only need to take $|{\bf P}|-(q-r'+N({\bf D}))$-element subset and there are $2(q-r'+N({\bf D}))$ elements remained. At the $j$-th step we have $(j-1)(q-r'+N({\bf D}))$ elements remained in the set $P_{j(r+1)} \times {\bf F}_q$. \\

When $(j-1)(q-r'+N({\bf D})) \geq N-N({\bf D})$, we add $j+1$ such codewords satisfying that\\
1) The intersection of any two of their supports has at most $N({\bf D})$ elements;\\
2) The intersection of any such support with the any image support has at most $r+1$ elements.\\

If $r+1 \leq N({\bf D})$, finally we add at least $[\frac{N}{r+1}]+\frac{N}{[\frac{N-N({\bf D})}{q-r'+N({\bf D})}]+1}$ such weight $N$ codewords.\\

{\bf Theorem  3.2.} {\em Suppose $|{\bf Y(F_q)-D}|=N=rq+r'$ and $q >r'\geq 0$. We assume $r+1 \leq N({\bf D})$. Then  $A(q \cdot N, 2(N-N({\bf D})), N) \geq  q^{dim(L({\bf D}))} +[\frac{N}{r+1}]+\frac{N}{[\frac{N-N({\bf D})}{q-r'+N({\bf D})}]+1}$.}\\

\section{Examples: curves}

{\bf Elliptic curves.} Let ${\bf E}$ be an elliptic curve over ${\bf F}_q$ with $N$ rational points ${\bf P}=\{P_1,...,P_N\}$. Suppose $N=rq+r'$ as in Theorem 3.1. We have $A(qN, 2(N-s), N) \geq q^s+\frac{N}{[\frac{N-s}{q-r'+s}]+1}$  if there is a degree $s$ rational divisor ${\bf G}$ whose support satisfying $supp{\bf G} \cap {\bf P}$ is empty, $1<s<N$ and $r+1 \leq s$.\\

{\bf Example 1.} The elliptic curve $y^2=x^3-2x-3$ defined over ${\bf F}_7$ has $10$ rational points $(3, 2), (2, 6), (4, 2), (0, 5), (5, 0), (0, 2), (4, 5 ), (2, 1), (3, 5)$ and the zero element (infinity point). It is clear it has a degree $s$ rational point. Thus we have $A(70, 20-2s, 10) \geq 7^s+[\frac{10}{2}]+\frac{5(4+s)}{7}$. When $s=2$, $A(70, 16, 10) \geq 59$ ($A(70, 16, 9)=49$ in \cite{table}). If only  $9$ rational points are used,  $A(63, 18-2s, 9) \geq 7^s+[\frac{9}{2}]+\frac{9(4+s)}{13}$ when $0<s<9$. Thus $A(63, 14, 9) \geq 57$ ($A(63, 14, 8) \geq 63$ in \cite{table}).\\

{\bf Example 2.} There is an elliptic curve over ${\bf F}_8$ with $14$ rational points (maximal curve, \cite{table2}). Thus we have $A(112, 28-2s, 14) \geq 8^s+[\frac{14}{2}]+7+\frac{7(2+s)}{8}$ for $1<s<14$, $A(104, 26-2s, 13) \geq 8^s+6+\frac{13(3+s)}{16}$ when $1<s<13$, $A(96, 24-2s, 12) \geq 8^s+6+\frac{3(4+s)}{4}$ when $1<s<12$, $A(88, 22-2s, 11) \geq 8^s+5+\frac{11(5+s)}{16}$ when $1<s<11$, and $A(80, 20-2s, 10) \geq 8^s+5+\frac{5(6+s)}{8}$ when $1<s<10$.\\

\begin{center}
{\bf Table 3} Explicit constant weight codes from EC\\
\bigskip
\begin{tabular}{||c|c||c||c||}\hline
Explicit codes& closest codes in \cite{table}&G-S bound & G bound\\ \hline
$A(80, 16, 10) \geq 74$&$A(80,16, 9) \geq 80$& $\leq1$&9.43 \\ \hline
$A(72, 14, 9) \geq 74$&$A(72, 14, 8) \geq 89$&$\leq1$&12.76 \\ \hline
$A(70, 16, 10) \geq 59$&$A(70, 16, 9) \geq 49$&$\leq1$&6.80 \\ \hline
$A(63, 14, 9) \geq 57$&$A(63, 14, 8) \geq 63$&$\leq1$& 9.07 \\ \hline
$A(36, 14, 9) \geq 23$& $A(36, 14, 8)=9$&$\leq 1$&2.51\\ \hline
$A(36, 12, 9) \geq 72$& $66 \geq A(36, 12, 8) \geq 45$&$\leq 1$& 7.45 \\ \hline
$A(36, 10, 9) \geq 265$& $A(36, 10, 8) \geq 216$&$\leq 1$&38.12\\ \hline
\end{tabular}
\end{center}

\begin{center}
{\bf Table 4} Explicit constant weight codes from EC\\
\bigskip
\begin{tabular}{||c|c||c||}\hline
Explicit codes& G-S bound& G bound  \\ \hline
$A(104, 22, 13) \geq 75$& $\leq1$& 4.85\\ \hline
$A(104, 20, 13) \geq 523$&$\leq1$&17.86\\ \hline
$A(104, 18, 13) \geq 4107$&$\leq1$&98.28\\ \hline
$A(104, 16, 13) \geq 32781$&$93$&810.42\\ \hline
$A(96, 20, 12) \geq 75$& $\leq1$&5.86\\ \hline
$A(96, 14, 12) \geq 52784$&$798.06$&1557.72\\ \hline
$A(88, 18, 11) \geq 73$&$\leq1$&7.30\\ \hline
$A(88, 16, 11) \geq 523$&$\leq1$&35.07\\ \hline
\end{tabular}
\end{center}

The above constant weight codes  are much better than the codes from the Graham-Sloane lower bound and Gilbert type lower bound.\\

{\bf Example 3. } Let ${\bf E}$ be the elliptic curve $y^2+y=x^3+x$ defined over ${\bf F}_{2^r}$ where $r \equiv 4$ $mod$ $8$. There are $N=2^r+2^{\frac{r}{2}+1}+1$ (see \cite{Washington}, Theorem 4.12) rational points on this curve. We have $A(2^{2r}+2^{\frac{3r}{2}+1}+2^r, 2(2^r+2^{\frac{r}{2}+1}-s), 2^r+2^{\frac{r}{2}+1}+1) \geq 2^{rs}+2^{r-1}+2^{\frac{r}{2}}+\frac{(2^r+2^{\frac{r}{2}+1}+1)(2^r-2^{\frac{r}{2}+1}-s)}{2^{r+1}}$ when $1<s<2^r+2^{\frac{r}{2}+1}$.\\

{\bf Hermitian curves.} The Hermitian curve over ${\bf F}_{q^2}$ is defined by $x^q+x=y^{q+1}$. It is well-known there are $N=q^3+1$ rational points. Thus $A(q^5, 2(q^3-s), q^3) \geq q^{2(s-\frac{q^2-q}{2}+1)}+q^2-1+\frac{q^3(q^2+s)}{q^3+q^2}$ when $q^2-q-2<s<q^3$ from Theorem 3.1.  Most of these explicit constructed binary constant weight codes are much better than the Gilbert type lower bound.\\

Just as the previous section expliciltly constructed binary constant weight codes from elliptic curves can be used to improve Gilbert and Graham-Sloane lower bounds.\\

\section{Examples: higher dimension case}

{\bf 5.1. Projective spaces.} In Theorem 3.2, we take ${\bf P}_{{\bf F}_q}^n$  and ${\bf D}=r{\bf H}$ the divisor. Then  $A(q^{n+1}, 2(q^n-rq^{n-1}-q^{n-2}-\cdots-q-1), q^n) \geq q^{\displaystyle{n+r \choose r}}+q-1+\frac{q+rq^{n-1}+q^{n-2}+\cdots+q+1}{2}$ from the Segre-Serre-Sorensen bound (\cite{HK}). In particular $A(q^3, 2(q^2-rq-1), q^2) \geq q^{\frac{(r+2)(r+1)}{2}}+q-1+\frac{q+rq+1}{1+1/q}$. We list some such explicit binary constant weight codes in Table 5. Except the second $A(64, 14, 16)=4107<4603$, all others  are much better than Gilbert type lower bound. Considering Gilbert lower bound is not constructive, our this code $A(64, 14, 16) \geq 4107$ is good.\\

\begin{center}
{\bf Table 5} Explicit constant weight codes from projective surface\\
\bigskip
\begin{tabular}{||c|c||}\hline
Explicit codes& Gilbert type bound \\ \hline
$A(64, 22, 16) \geq 72$& $6.31$\\ \hline
$A(64, 14, 16) \geq 4107 $&$4603.81$\\ \hline
$A(125, 38, 25) \geq $135&$5.05$\\ \hline
$A(125, 28, 25) \geq 15639$&$3015.31$\\ \hline
\end{tabular}
\end{center}

{\bf 5.2. Ruled Surface.} In Theorem 3.2 we take ${\bf X}={\bf P}_{{\bf F}_q}^1 \times {\bf P}_{{\bf F}_q}^1$. The set of ${\bf F}_q$ rational points on ${\bf P}_{{\bf F}_q}^1 \times {\bf P}_{{\bf F}_q}^1$ is naturally the disjoint union of $(q+1)$ sets of ${\bf F}_q$ rational points on curves $p_i \times {\bf P}_{{\bf F}_q}^1$, where $p_i$, $i=1,...,q+1$ are $(q+1)$ rational points of ${\bf P}_{{\bf F}_q}^1$.  The divisor ${\bf D}$ of type $(d_1,d_2)$ consists of polynomials $f(x,y,z,w)$ which are homogeneous in $x,y$ with degree $d_1$ and are homogeneous in $z,w$ with degree $d_2$. This is a linear system with dimension $(d_1+1)(d_2+1)$. If $d_1+d_2 <q+1$, there are at most $-d_1d_2+d_1(q+1)+d_2(q+1)$ rational points on any member of this linear system (\cite{Hansen}). Then $A(q(q+1)^2, 2((q+1)^2+d_1d_2-(d_1+d_2)(q+1)), (q+1)^2) \geq q^{(d_1+1)(d_2+1)}+q-1+\frac{(q+1)^2(q-1+(d_1+d_2)(q+1)-d_1d_2)}{(q+1)^2+q-1}$ from Theorem 3.2. We list some binary constant weight codes in the following Table 6. Some of them are much better than Gilbert type lower bound.\\

\begin{center}
{\bf Table 6} Explicit constant weight codes from ruled surface\\
\bigskip
\begin{tabular}{||c|c||}\hline
Explicit codes& Gilbert type bound \\ \hline
$A(100, 24, 25) \geq 4111$& $1771.61$\\ \hline
$A(180, 40, 36) \geq 15647 $&$39467.85$\\ \hline
$A(180, 50, 36) \geq 643$&$38.31$\\ \hline
$A(448, 84, 64) \geq 117681$&$616907.85$\\ \hline
\end{tabular}
\end{center}

{\bf 5.3. Toric surfaces.}  Algebraic geometric codes from toric surfaces have been studied in \cite{JHansen}. In this section we give some exlpicit binary constant weight codes from some toric surfacse in \cite{JHansen}.\\

Let ${\bf Z}^2 \subset {\bf R}^2$ be the set of all integral points . We denote $\theta$ a primitive element of the finite field ${\bf F}_q$. For any integral point ${\bf m}=(m_1, m_2) \in {\bf Z}^2$ we have a function $e({\bf m}): {\bf F}_q^{*} \times {\bf F}_q^{*} \rightarrow {\bf F}_q$ defined as $e({\bf m})(\theta^i,\theta^j)=\theta^{m_1i+m_2j}$ for $i=0,1,...,q-1$ and $j=0,1,...,q-1$. Let $\Delta \subset {\bf R}^2$ be a convex polyhedron with vertices in ${\bf Z}^2$ and $L(\Delta)$ be the function space over ${\bf F}_q$ spanned by these functions $e({\bf m})$ where ${\bf m}$ takes over all integral points in $\Delta$. In the following cases of convex polyhedrons these functions are linearly independent from \cite{JHansen}.\\

The following three cases as in  the main results Theorem 1, 2, 3 of \cite{JHansen} are considered.\\
1) $\Delta$ is the convex polytope with the vertices $(0,0), (d,0), (0,d)$ where $d$ is a positive integer satisfying $d <q-1$;\\
2) $\Delta$ is the convex polytope with the vertices $(0,0), (d,0), (d,e+rd), (0,e)$ where $d, r, e$ are positive integers satisfying $d<q-1$, $e<q-1$ and $e+rd<q-1$;\\
3) $\Delta$ is the convex polytope with the vertices $(0,0), (d,0), (0,2d)$ where $d$ is a positive integer satisfying $2d <q-1$;\\

For each function $f \in L(\Delta)$ we have a length $q \times (q-1)^2$ codeword ${\bf v}(f)=(f_{(a,b)})$ where $(a,b) \in {\bf F}_q^{*} \times {\bf F}_q^{*} \times {\bf F}_q$ defined as follows. $f_{(a,b)}=0$ if $f(a) \neq b$ and $f_{(a,b)}=1$ if $f(a)=b$. The Hamming weight of this codeword is exactly $(q-1)^2$. Then there are $q^{dim(L(\Delta))}$ such weight $(q-1)^2$ codewords. 
 We have the following result from the main results Theorem 1, 2, 3 of \cite{JHansen} and Theorem 3.2.\\

{\bf Proposition 5.1.} {\em In the above cases we have\\
1) $A(q(q-1)^2, 2((q-1)^2-d(q-1)), (q-1)^2) \geq q^{\frac{(d+1)(d+2)}{2}}+q-1+\frac{(q-1)^2(d+1)}{q}$ in the case 1);\\
2) $A(q(q-1)^2, 2((q-1)^2-min\{(d+e)(q-1)-de, (e+rd)(q-1)\}), (q-1)^2) \geq q^{(d+1)(e+1)+\frac{rd(d+1)}{2}}+q-1+\frac{(q-1)(q-1+ min\{(d+e)(q-1)-de, (e+rd)(q-1)\})}{q}$ in the case 2);\\
3) $A(q(q-1)^2, 2((q-1)^2-2d(q-1)), (q-1)^2) \geq q^{d^2+2d+1}+q-1+\frac{(q-1)^2(2d+1)}{q}$ in the case 3).}\\

In the following table we list some explicit binary constant weight codes from toric surfacse. They are much better than Gilbert type lower bound.\\

\begin{center}
{\bf Table 7} Explicit constant weight codes from toric surfaces\\
\bigskip
\begin{tabular}{||c|c||}\hline
Explicit codes& Gilbert type bound \\ \hline
$A(80, 24, 16) \geq 136$& $5.57$\\ \hline
$A(80, 18, 16) \geq 3138$&$416.62$\\ \hline
$A(80, 12, 16) \geq 1953141$&$781764.18$\\ \hline
\end{tabular}
\end{center}

\section{Summary}
Explicit binary constant weight codes have been constructed from Reed-Solomon codes. This is a strict improvement on the previous works in \cite{DeVore,Ericson}. Examples of nice binary constant weight codes have been given. Some of new better binary constant weight codes have been explicitly constructed. The parameters of most of our explicit binary constant weight codes are much better than the Gilbert type lower bound and Graham-Sloane lower bound. Asmptotically our explicit binary constant weight codes even from Reed-Solomon codes have parameters better than the non-explicit Gilbert and Graham-Sloane lower bounds in some range of parameters. We also give an extension to algebraic geometric codes and many good binary constant weight codes are explicitly constructed.

\bibliographystyle{amsplain}

\begin{thebibliography}{10}
\bibitem{Vardy} E. Agrell, A. Vardy, and K. Zeger, Upper bounds for constant-weight codes, IEEE Trans. Inf. Theory, vol. 46 (2000), no. 7, 2373–2395.

\bibitem{Brouwer} A. E. Brouwer and T. Etzion, Some new constant weight codes, Advances in Mathematics of Communications, vol. 5 (2011), 417-424.

\bibitem{CHT}  R. Calderbank, M. A. Herro, and V. Telang, A multilevel approach to the design of DC-free line codes, IEEE Trans. Inform. Theory, vol. 35 (1989) 579-583.\\

\bibitem{Chee} Y. M. Chee, C. Xing and S. Z. Ling, New constant-weight codes from propagation rules, IEEE Transactions on Information Theory, vol. 56 (2010), 1596-1599.\\
\bibitem{DeVore} R. DeVore, Deterministic constructions of compressed sensing matrices, Journal of Complexity, Vol.23 (2007), no.46,  918-925.

\bibitem{Ericson} T. Ericson and V. A. Zinoviev, An improvement of the Gilbert bound for constant weight codes, IEEE Transactions on Information Theory, vol.33 (1987), 721-723.\\

\bibitem{EtzionVardy} T. Etzion and A. Vardy, A new construction for constant weight codes, ArXiv:1004.1503v3.\\



\bibitem{Graham} R. L. Graham and N. J. A. Sloane, Lower bounds for constant weight codes, IEEE Trans. Inf. Theory, vol.26 (1980), no. 1, pp. 37-43.
\bibitem{Massey} N. Q. A, L. Gyorfri and J. L. Massey, Constructions of binary constant- weight cyclic codes and cyclically permutable codes, IEEE Trans. Inform. Theory, vol. 38 (1992), 940-949.\\
\bibitem{Immink} K. A. Immink, Coding Techniques for Digital Recorders. London: Prentice-Hall, 1991.
\bibitem{JHansen} J. P. Hansen, Toric surfaces and codes, techniques and examples, Coding theory, cryptography and related areas, ed. J. Bachmann et al., Springer, 2000.
\bibitem{Hansen} S. H. Hansen, Error-correcting codes over higher dimensional varieties,  Finite Fields and Their Applications,Vol. 7 (2001), 530-552.
\bibitem{Hartshorne} R. Hartshorne, Algebraic geometry, Springer-Verlag, 1977.
\bibitem{HK} M.Homma and S.J.Kim, An elementary bound for the number of rational points of a hypersurface over finite fields, Finite Fields and Their Applications, Vol.20 (2014), 76-83.

\bibitem{Johnson} S. M. Johnson, A new upper bound for error-correcting codes, IRE Trans. Inform. Theory, vol. IT-8 (1962), pp. 203-207.

\bibitem{TV} M. A. Tsfasman and S. G. Vladut, Algebraic-geometric codes, Dordrecht, Kluwer, 1991.


\bibitem{Washington} L.C.Washington, Elliptic Curves: Number Theory and Cryptography, Discr. Math. Appl.(series), 2nd ed. Boca Raton, FL: CRC Press, 2008.

\bibitem{XingLing} C. Xing and J. Ling, A construction of binary constant weight codes from algebraic curves over finite fields, IEEE Transactions on Information Theory, vol.51(2005), 3674-3678.


\bibitem{table} http://www.win.tue.nl/~aeb/Andw.html


\bibitem{table2} http://gerard.vdgeer.net/tables-mathcomp21.pdf


\end{thebibliography}

\end{document}